\documentclass[showpacs,showkeys,pra,12pt,tightenlines]{revtex4}
\usepackage{graphicx}

\begin{document}

\title{Quantumness tests and witnesses \\in the tomographic--probability representation}

\author{S. N. Filippov$^1$}
\email{filippovsn@gmail.com}

\author{V. I. Man'ko$^2$}
\email{manko@sci.lebedev.ru}

\affiliation{$^1$ Moscow Institute of Physics and Technology
(State University) \\Institutskii per. 9, Dolgoprudnyi, Moscow
Region 141700, Russia
\\$^2$ P. N. Lebedev Physical Institute, Russian Academy of Sciences \\Leninskii Prospect 53, Moscow 119991, Russia}

\begin{abstract}
In view of the tomographic--probability representation of quantum
states, we reconsider the approach to quantumness tests of a
single system developed in [Alicki and Van Ryn 2008 {\it J. Phys.
A: Math. Theor.} {\bf 41} 062001]. For qubits we introduce a
general family of quantumness witnesses which are operators
depending on an extra parameter. Spin tomogram and dual spin
tomographic symbols are used to study qubit examples and the test
inequalities which are shown to satisfy simple relations within
the framework of the standard probability theory.
\end{abstract}



\maketitle

\section{\label{introduction}Introduction}

A boundary between quantum and classical worlds is rather
difficult to draw precisely, while the problem of their
distinguishing is getting more important than it was hardly ever
before. This problem is particularly important for the large-scale
quantum computer to be constructed because it must be a
macroscopic object and exhibit quantum properties at the same
time. The investigations in this field of science are also
stimulated by the attempts to describe quantum mechanical
phenomena by different kinds of hidden variables models. Though
these problems have been paid a great attention for many years the
common point of view has not been achieved. The discussions came
up with a bang after a recent proposal
\cite{alicki-small,alicki-large,alicki-JJ,alicki-critic} of a
simple test of checking whether it is possible or not to describe
a given set of experimental data by a classical probabilistic
model.

In quantum mechanics the state of a system can be identified with
a fair probability called tomographic probability distribution or
state tomogram (see, e.g., the review \cite{sudarshan-2008}). The
probability representation of quantum states with continuous
variables (position, momentum) was introduced in
\cite{tombesi-manko} and that of states with discrete variables
(spin, qudits) was introduced in \cite{dodonovPLA,oman'ko-jetp}.
In the probability representation of quantum mechanics the
relation between classical and quantum systems behavior can be
studied using the same notion of states expressed in terms of the
probability distribution (tomogram) in both classical and quantum
domains \cite{oman'ko-97,mendes-physica}. The quantum (or
classical) nature of a system state can be clarified if in the
system the uncertainty relations
\cite{heisenberg,schrodinger,robertson30,robertson34} for
conjugate variables are fulfilled. Also, if the state is quantum,
it can be similar to a classical one and there are some studies of
the classicality (see, e.g., \cite{shchukin-vogel}) based on
properties of the diagonal representation of density operator
\cite{sudarshan-63} (or \textsl{P}-function \cite{glauber}).

In the works \cite{alicki-small,alicki-large}, quantum or
classical properties of a system state were associated with
measuring some specific observables such that there exist certain
inequalities which hold true in classical domain and are violated
in quantum domain. Violation of the inequalities is considered as
a quantumness witness of the system state. In this sense, the
criterion \cite{alicki-small,alicki-large} is similar in its
spirit to the Bell inequalities \cite{bell,chsh}. The Bell
inequalities were studied by means of the tomographic--probability
representation in
\cite{lupo1,lupo2,andreev-shchukin,filipp-qubit-portrait}. The aim
of our work is to consider the inequalities introduced in
\cite{alicki-small,alicki-large} and their properties in classical
and quantum domains within the framework of the probability
representation of quantum states. We suppose that such a procedure
is necessary while dealing with the quantum probabilistic model
based on tomograms.

\bigskip

The paper is organized as follows.

In Sec. \ref{spin-tomogram}, we are aimed at recalling the
tomographic representation of qubit states and observables by
employing both ordinary and dual tomographic symbols. In Sec.
\ref{section-quantumness-tests}, the quantumness test is discussed
within the framework of classical and quantum probability
descriptions. In Sec. \ref{example}, we present a family of
observables which can be used to detect quantumness of an
arbitrary generally mixed state $\hat{\rho} \neq \hat{1}/n$ of a
single system. Here we also predict what kind of experiment one
should make to test the quantumness of a state specified by its
tomogram. In Sec. \ref{conclusions}, conclusions and prospects are
presented.

\section{\label{spin-tomogram}States and observables in the tomographic--probability representation}

Apart from being described by the conventional density operator
$\hat{\rho}$, the state of a qubit is determined thoroughly by its
spin tomogram. The probability distribution function (spin
tomogram) $w(m,u)$ is nothing else but the probability to obtain
$m$ ($m = \pm \textstyle{1 \over 2}$) as spin projection on the
direction given by the unitary $2\times 2$ matrix $u$. This matrix
can be considered as a matrix of an irreducible representation of
the rotation group depending on two Euler angles determining the
direction of quantization (point on the Bloch sphere). The
relation between $\hat{\rho}$ and $w(m,u)$ reads

\begin{equation}
\label{tomogram-definition} w({\bf{x}}) \equiv w(m,u) = \langle m
| u \hat{\rho} u^{\dag} | m \rangle \nonumber = {\rm Tr} \left(
\hat{\rho} \ u^{\dag} | m \rangle \langle m| u \right) = {\rm Tr}
\left( \hat{\rho} \hat{U}({\bf{x}}) \right),
\end{equation}

\noindent where the operator $\hat{U}({\bf{x}}) = u^{\dag} | m
\rangle \langle m| u$ is called the dequantizer operator and
${\bf{x}}$ is assigned to the set of parameters $(m,u)$. The
general procedure to use dequantizer operators was discussed in
the context of star-product quantization schiemes in
\cite{marmoJPA,oman'ko-star-product}. The explicit form of the
dequantizer $\hat{U}({\bf{x}})$ can be obtained readily by
exploiting the matrix $u$ expressed in terms of the Euler angles
$\alpha$, $\beta$, $\gamma$:

\begin{equation}
\label{u-matrix}
u = \left(%
\begin{array}{cc}
  \cos(\beta / 2) \ {\rm e}^{-{\rm i} (\alpha + \gamma)/2} & -\sin(\beta / 2) \ {\rm e}^{{\rm i} (\alpha - \gamma)/2} \\
  \sin(\beta / 2) \ {\rm e}^{-{\rm i} (\alpha - \gamma)/2} & \cos(\beta / 2) \ {\rm e}^{{\rm i} (\alpha + \gamma)/2} \\
\end{array}%
\right).
\end{equation}

\noindent Therefore, taking advantage of $| m \rangle \langle m |
= \textstyle{1 \over 2}{\hat{I}} + m \hat{\sigma}_{z}$, where
${\hat{I}}$ is the $2 \times 2$ identity matrix and
$\hat{\sigma}_{z}$ is the third Pauli matrix, one can write

\begin{equation}
\hat{U}({\bf{x}}) = \textstyle{1 \over 2}\hat{I} + m
\hat{F}(\alpha,\beta),
\end{equation}

\noindent where ${\bf{x}}=(m,\alpha,\beta,\gamma)$ and the matrix
$\hat{F}(\alpha,\beta)$ has the following form

\begin{equation}
\hat{F}(\alpha,\beta) = \left(%
\begin{array}{cc}
  \cos\beta & -{\rm e}^{{\rm i}\alpha}\sin\beta \\
  -{\rm e}^{-{\rm i}\alpha}\sin\beta & -\cos\beta \\
\end{array}%
\right).
\end{equation}

If given the spin tomogram $w({\bf{x}})$ it is possible to
reconstruct the density operator $\hat{\rho}$ \cite{castanos}.
This reconstruction was shown to take the simple form

\begin{equation}
\hat{\rho} = \int { w({\bf{x}}) \hat{D}({\bf{x}}) {\rm d} {\bf{x}}
},
\end{equation}

\noindent where the integration implies

\begin{equation}
\int {\rm d} {\bf{x}} = \sum \limits_{m=-1/2}^{1/2} {
\int\limits_{0}^{2\pi} {\rm d}\alpha \int\limits_{0}^{\pi}
\sin\beta {\rm d}\beta \int\limits_{0}^{2\pi} {\rm d}\gamma},
\end{equation}

\noindent and the quantizer operator $\hat{D}({\bf{x}})$ is
defined by the formula

\begin{equation}
\hat{D}({\bf{x}}) = \frac{1}{8\pi^{2}} \left( \frac{1}{2}\hat{I} +
3m \hat{F}(\alpha,\beta) \right).
\end{equation}

In quantum mechanics any observable $A$ is identified with a
Hermitian operator $\hat{A}$. By analogy with the density operator
$\hat{\rho}$ one can introduce the tomographic symbol
$w_{A}({\bf{x}})$ of the operator $\hat{A}$. Just in the same way
we write

\begin{equation}
w_{A}({\bf{x}}) = {\rm Tr} \left( \hat{A} \hat{U}({\bf{x}})
\right),\qquad \hat{A} = \int  w_{A}({\bf{x}}) \hat{D}({\bf{x}})
{\rm d} {\bf{x}}.
\end{equation}

It is worth noting that both quantizer and dequantizer are
operators depending on the set of parameters ${\bf{x}}$ so it
seems possible to swap quantizer with the dequantizer.
Substituting the quantizer operator for the dequantizer one and
visa versa leads to a so-called dual tomographic symbol
$w_{A}^{d}({\bf{x}})$ \cite{oman'ko-star-product,patr} satisfying
the following relations:

\begin{equation}
w_{A}^{d}({\bf{x}}) = {\rm Tr} \left( \hat{A} \hat{D}({\bf{x}})
\right),\qquad \hat{A} = \int  w_{A}^{d}({\bf{x}})
\hat{U}({\bf{x}}) {\rm d} {\bf{x}}.
\end{equation}

The dual symbol in the tomographic--probability representation
turned out to provide the function introduced in \cite{oman'ko-97}
which after averaging with a tomogram yields the mean value of the
observable. Let us now express the average value of the observable
$A$ by means of ordinary and dual tomographic symbols. Indeed, the
mean value of $A$ reads

\begin{equation}
{\rm Tr}\left(\hat{\rho}\hat{A}\right) = {\rm Tr} \int {\rm d}
{\bf{x}} \ w({\bf{x}}) \hat{D}({\bf{x}}) \hat{A} = \int {\rm d}
{\bf{x}} \ w({\bf{x}}) {\rm Tr} \left( \hat{A} \hat{D}({\bf{x}})
\right) \nonumber = \int w({\bf{x}}) w_{A}^{d}({\bf{x}}) {\rm d}
{\bf{x}}.
\end{equation}

\noindent The formula obtained can be checked immediately for a
general case of the density operator $\hat{\rho}$ and the
observable $A$:

\begin{equation}
\label{matrix-form-rho-A}
\hat{\rho} = \left(%
\begin{array}{cc}
  \rho_{11} & \rho_{12}{\rm e}^{{\rm i}\zeta} \\
  \rho_{12}{\rm e}^{-{\rm i}\zeta} & \rho_{22} \\
\end{array}%
\right),  \qquad \hat{A} = \left(%
\begin{array}{cc}
  A_{11} & A_{12}{\rm e}^{{\rm i}\eta} \\
  A_{12}{\rm e}^{-{\rm i}\eta} & A_{22} \\
\end{array}%
\right),
\end{equation}

\noindent where $\rho_{ij}$, $A_{ij}$, $\zeta$, $\eta$ are real
numbers. Then the spin tomogram $w({\bf{x}})$ is

\begin{equation}
w({\bf{x}}) = \textstyle{1 \over 2} + m
(\rho_{11}-\rho_{22})\cos\beta -
2m\rho_{12}\cos(\zeta-\alpha)\sin\beta
\end{equation}

\noindent and the dual tomographic symbol $w_{A}^{d}({\bf{x}})$
reads

\begin{equation}
w_{A}^{d}({\bf{x}}) =  \frac{1}{8\pi^{2}} \bigg[ \frac{1}{2}
(A_{11}+A_{22}) + 3m (A_{11}-A_{22}) \cos\beta - 6 m A_{12}
\cos(\eta-\alpha) \sin\beta \bigg].
\end{equation}

\noindent The direct calculation yields

\begin{equation}
\int w({\bf{x}}) w_{A}^{d}({\bf{x}}) {\rm d} {\bf{x}} =
\textstyle{1 \over 2}(A_{11}+A_{22})  + \textstyle{1 \over 2}
(\rho_{11}-\rho_{22}) (A_{11}-A_{22})  +
2\rho_{12}A_{12}\cos(\zeta-\eta)
\end{equation}

\noindent that coincides totally with the quantity ${\rm
Tr}(\hat{\rho}\hat{A})$ computed by using the matrix form of
operators (\ref{matrix-form-rho-A}).

Let us now express the average value of the observable $A$ in
terms of its possible outcomes $A_{\uparrow}$ and $A_{\downarrow}$
measured through separate experiments. The numbers $A_{\uparrow}$
and $A_{\downarrow}$ are nothing else but eigenvalues of the
operator $\hat{A}$. Consequently there exists a unitary matrix
$u_{A}$ such that the matrix $\hat{A}$ can be factorized as
follows

\begin{equation}
\hat{A} = u_{A}^{\dag}\left(%
\begin{array}{cc}
  A_{\uparrow} & 0 \\
  0 & A_{\downarrow} \\
\end{array}%
\right)u_{A}.
\end{equation}

\noindent The matrix $u_{A}^{\dag}$ is composed of two columns
which are eigenvectors of the operator $\hat{A}$ corresponding to
eigenvalues $A_{\uparrow}$ and $A_{\downarrow}$, respectively. In
order to specify the matrix $u_{A}$ one can substitute the Euler
angles $(\phi,\theta,\varphi)$ for $(\alpha,\beta,\gamma)$ in
formula (\ref{u-matrix}), i.e., $u_{A}=u(\phi,\theta,\varphi)$.
Then the dual tomographic symbol of the operator $\hat{A}$ takes
the form

\begin{equation}
w_{A}^{d}({\bf{x}}) = \frac{1}{8\pi^{2}} \bigg\{
\frac{1}{2}(A_{\uparrow}+A_{\downarrow}) +
3m(A_{\uparrow}-A_{\downarrow}) \times\left[\cos\beta \cos\theta +
\cos(\alpha-\phi) \sin\beta \sin\theta\right] \bigg\}.
\end{equation}

\noindent Now, when $w_{A}^{d}({\bf{x}})$ is known, it is not
impossible to evaluate the integral

\begin{eqnarray}
\int w({\bf{x}}) w_{A}^{d}({\bf{x}}) {\rm d} {\bf{x}} && =
\textstyle{1 \over 2}(A_{\uparrow}+A_{\downarrow}) + \textstyle{1
\over 2} (A_{\uparrow}-A_{\downarrow}) (\rho_{11}-\rho_{22})
\cos\theta -
(A_{\uparrow}-A_{\downarrow})\rho_{12}\cos(\zeta-\phi) \nonumber
\\
&& = A_{\uparrow} \left[ \textstyle{1 \over 2} + \textstyle{1
\over 2}(\rho_{11}-\rho_{22})\cos\theta - \rho_{12}\sin\theta
\cos(\zeta-\phi) \right] \nonumber \\ && \qquad + A_{\downarrow}
\left[ \textstyle{1 \over 2} - \textstyle{1 \over
2}(\rho_{11}-\rho_{22})\cos\theta + \rho_{12}\sin\theta
\cos(\zeta-\phi) \right] \nonumber
\\ && = w(+\textstyle{1 \over 2},\phi,\theta,\varphi)A_{\uparrow} +
w(-\textstyle{1 \over 2},\phi,\theta,\varphi)A_{\downarrow}
\nonumber \\ && = w_{\uparrow}(u_{A})A_{\uparrow} +
w_{\downarrow}(u_{A})A_{\downarrow},
\end{eqnarray}

\noindent that gives the average value of the observable $A$,
i.e., the value of quantity ${\rm Tr}(\hat{\rho}\hat{A})$. Here we
denoted $w(m=+\textstyle{1 \over 2},u)$ and $w(m=-\textstyle{1
\over 2},u)$ by $w_{\uparrow}(u)$ and $w_{\downarrow}(u)$,
respectively. One cannot help mentioning that the same result is
achieved by using the definition of the spin tomogram
(\ref{tomogram-definition}):

\begin{eqnarray}
{\rm Tr}\left(\hat{\rho}\hat{A}\right) && =
{\rm Tr}\left[\hat{\rho}u_{A}^{\dag}\left(%
\begin{array}{cc}
  A_{\uparrow} & 0 \\
  0 & A_{\downarrow} \\
\end{array}%
\right)u_{A}\right] = {\rm Tr}\left[u_{A}\hat{\rho}u_{A}^{\dag}\left(%
\begin{array}{cc}
  A_{\uparrow} & 0 \\
  0 & A_{\downarrow} \\
\end{array}%
\right)\right] \nonumber \\
&& = {\rm Tr}\left[ \left(%
\begin{array}{cc}
  \langle \textstyle{1 \over 2} | u_{A}\hat{\rho}u_{A}^{\dag} | \textstyle{1 \over 2} \rangle & \langle \textstyle{1 \over 2} | u_{A}\hat{\rho}u_{A}^{\dag} | -\textstyle{1 \over 2} \rangle \\
  \langle -\textstyle{1 \over 2} | u_{A}\hat{\rho}u_{A}^{\dag} | \textstyle{1 \over 2} \rangle & \langle -\textstyle{1 \over 2} | u_{A}\hat{\rho}u_{A}^{\dag} | -\textstyle{1 \over 2} \rangle \\
\end{array}%
\right) \left(%
\begin{array}{cc}
  A_{\uparrow} & 0 \\
  0 & A_{\downarrow} \\
\end{array}%
\right)\right] \nonumber \\
&& =  \langle \textstyle{1 \over 2} | u_{A}\hat{\rho}u_{A}^{\dag}
| \textstyle{1 \over 2} \rangle A_{\uparrow} + \langle
-\textstyle{1 \over 2} | u_{A}\hat{\rho}u_{A}^{\dag} |
-\textstyle{1 \over 2} \rangle A_{\downarrow} \nonumber \\
&& = w_{\uparrow}(u_{A})A_{\uparrow} +
w_{\downarrow}(u_{A})A_{\downarrow}.
\end{eqnarray}

\section{\label{section-quantumness-tests} Quantumness tests}

In this section, we are going to answer the question whether it is
possible or not to describe the system involved by a classical
probabilistic model. The negative reply indicates straightway the
quantumness of the system in question. To start, let us consider
the case of a single qubit, and then discuss a generalization for
an arbitrary qudit system.

In the classical statistical model of a single qubit system the
observable $A$ is associated with a set of real numbers
$(A_{\uparrow},A_{\downarrow})$, where the numbers $A_{\uparrow}$
and $A_{\downarrow}$ are nothing else but possible outcomes of the
measurement of $A$. Moreover, the observable $A^{2}$ has possible
outcomes of the form $(A_{\uparrow}^{2},A_{\downarrow}^{2})$,
i.e., is in strong relation to $A$. The states form a simplex of
probability distributions $(p_{\uparrow}, p_{\downarrow})$, with
$0 \le p_{\uparrow}$, $0 \le p_{\downarrow}$, and $p_{\uparrow} +
p_{\downarrow} = 1$ (such a simplex is a geometrical treatment of
all possible classical states; geometric interpretation of quantum
states is reviewed in \cite{filip-geometrical}). Then the
expectation values of observables $A$ and $A^{2}$ read

\begin{equation}
\langle A \rangle _{cl} = p_{\uparrow}A_{\uparrow} +
p_{\downarrow}A_{\downarrow},\qquad \langle A^{2} \rangle _{cl} =
p_{\uparrow}A_{\uparrow}^{2} + p_{\downarrow}A_{\downarrow}^{2}.
\end{equation}

It is worth noting that such a classical system is equivalent to
the quantum one, with both the density operator and the operator
$\hat{A}_{cl}$ being of the diagonal form

\begin{equation}
\hat{\rho}_{cl} = \left(%
\begin{array}{cc}
  p_{\uparrow} & 0 \\
  0 & p_{\downarrow} \\
\end{array}%
\right), \qquad \hat{A}_{cl} = \left(%
\begin{array}{cc}
  A_{\uparrow} & 0 \\
  0 & A_{\downarrow} \\
\end{array}%
\right).
\end{equation}

Suppose one has two nonnegative observables $A$ and $B$ such that
inequality $\langle A \rangle _{cl} \le \langle B \rangle _{cl}$
holds true for all states $(p_{\uparrow}, p_{\downarrow})$. That
implies the following relations: $0 \le A_{\uparrow} \le
B_{\uparrow}$ and $0 \le A_{\downarrow} \le B_{\downarrow}$. If
this is the case, the average value of observable $A^{2}$ is
necessarily less or equal than the average value of $B^{2}$.
Indeed,

\begin{equation}
\langle A^{2} \rangle _{cl} = p_{\uparrow}A_{\uparrow}^{2} +
p_{\downarrow}A_{\downarrow}^{2} \le p_{\uparrow}B_{\uparrow}^{2}
+ p_{\downarrow}B_{\downarrow}^{2} = \langle B^{2} \rangle _{cl}.
\end{equation}

\noindent The mathematical aspect of the classical statistical
model is briefly expressed by the implication

\begin{equation}
\label{classical-inequalities}
{\rm Tr} \left[ \left(%
\begin{array}{cc}
  p_{\uparrow} & p_{\uparrow} \\
  p_{\downarrow} & p_{\downarrow} \\
\end{array}%
\right) \left(%
\begin{array}{cc}
  A_{\uparrow} & A_{\downarrow} \\
  -B_{\uparrow} & -B_{\downarrow} \\
\end{array}%
\right) \right] \le 0 \qquad \Rightarrow \qquad  {\rm Tr} \left[ \left(%
\begin{array}{cc}
  p_{\uparrow} & p_{\uparrow} \\
  p_{\downarrow} & p_{\downarrow} \\
\end{array}%
\right) \left(%
\begin{array}{cc}
  A_{\uparrow}^{2} & A_{\downarrow}^{2} \\
  -B_{\uparrow}^{2} & -B_{\downarrow}^{2} \\
\end{array}%
\right) \right] \le 0,
\end{equation}

\noindent where we introduced a stochastic matrix with the matrix
elements $p_{\uparrow}$ and $p_{\downarrow}$.

In the quantum statistical model the observable $A$ is associated
with the operator $\hat{A}$, which in its turn corresponds to the
Hermitian matrix whose eigenvalues $A_{\uparrow}$ and
$A_{\downarrow}$ give possible outcomes of the measurement of $A$.
States are identified with density operators $\hat{\rho}$, which
are positive Hermitian matrices with the trace equal to unity. By
using the tomographic representation of quantum states and
observables developed in previous section, one can write the
average values of observables $A$ and $A^{2}$ as follows:

\begin{eqnarray}
\langle A \rangle _{q} = {\rm Tr}\left(\hat{\rho}\hat{A}\right) =
w_{\uparrow}(u_{A})A_{\uparrow} +
w_{\downarrow}(u_{A})A_{\downarrow}, \\
\langle A^{2} \rangle _{q} = {\rm
Tr}\left(\hat{\rho}\hat{A}^{2}\right) =
w_{\uparrow}(u_{A})A_{\uparrow}^{2} +
w_{\downarrow}(u_{A})A_{\downarrow}^{2}.
\end{eqnarray}

In a way analogues to the classical case, we consider two
nonnegative operators $\hat{A}$ and $\hat{B}$ (i.e., having
nonnegative eigenvalues
$A_{\uparrow},A_{\downarrow},B_{\uparrow},B_{\downarrow}\ge 0$)
such that the residual operator $\hat{B}-\hat{A}$ is nonnegative
as well. The last requirement can be rewritten in the form of the
condition

\begin{equation}
\label{quantum-first-moment}
{\rm Tr} \left[ \left(%
\begin{array}{cc}
  w_{\uparrow}(u_{A}) & w_{\uparrow}(u_{B}) \\
  w_{\downarrow}(u_{A}) & w_{\downarrow}(u_{B}) \\
\end{array}%
\right) \left(%
\begin{array}{cc}
  A_{\uparrow} & A_{\downarrow} \\
  -B_{\uparrow} & -B_{\downarrow} \\
\end{array}%
\right) \right] \le 0
\end{equation}

\noindent that must be valid for \emph{all} tomograms $w(m,u)$. In
(\ref{quantum-first-moment}), we introduced a stochastic matrix
with the matrix elements $w_{\uparrow}(u_{A,B})$ and
$w_{\downarrow}(u_{A,B})$. In contrast to the classical case, even
under these circumstances the average value of the operator
$\hat{B}^{2}-\hat{A}^{2}$ does not have to be nonnegative.
Therefore, there can exist the tomogram $\widetilde{w}(m,u)$ such
that

\begin{equation}
\label{violation}
{\rm Tr} \left[ \left(%
\begin{array}{cc}
  \widetilde{w}_{\uparrow}(u_{A}) & \widetilde{w}_{\uparrow}(u_{B}) \\
  \widetilde{w}_{\downarrow}(u_{A}) & \widetilde{w}_{\downarrow}(u_{B}) \\
\end{array}%
\right) \left(%
\begin{array}{cc}
  A_{\uparrow}^{2} & A_{\downarrow}^{2} \\
  -B_{\uparrow}^{2} & -B_{\downarrow}^{2} \\
\end{array}%
\right) \right] > 0.
\end{equation}

The difference between classical and quantum behaviour of the
observable $A^{2}-B^{2}$ serves as the basis for a simple
quantumness test.

\bigskip

\textbf{Quantumness Test}. Two observables $A$ and $B$ are found
which averaged values satisfy the inequality $0 \le \langle A
\rangle \le \langle B \rangle$ for all experimentally accessible
states, but for a certain state the condition $\langle A^{2}
\rangle \le \langle B^{2} \rangle$ is violated, then the state
involved cannot be described by a classical probabilistic model
and is surely to be quantum.

This is the reason why the operator $\hat{B}^{2}-\hat{A}^{2}$ was
proposed to call quantumness witness \cite{alicki-large}.

The comparison of the expressions (\ref{classical-inequalities})
and (\ref{violation}) shows that despite both classical and
quantum cases can be treated probabilistically the difference
between them occurs due to the classical state being associated
with numbers $(p_{\uparrow},p_{\downarrow})$ while the quantum one
being identified with the functions $w_{\uparrow}(u)$ and
$w_{\downarrow}(u)$ taking on the different values at the unitary
group elements $u=u_{A}$ and $u=u_{B}$.

The generalization of this approach for a qudit state given by its
$n \times n$ density matrix $\hat{\rho} \neq \hat{1}/n$ is evident
while dealing with operators $\hat{A}$, $\hat{B}$ having an effect
on the two-dimensional subspace spanned by eigenvectors
corresponding to two different eigenvalues of $\hat{\rho}$.

\section{\label{example}Quantumness Witnesses}

The quantumness test described above requires a certain
quantumness witness to be found. In the earlier work
\cite{alicki-large}, the abundance of quantumness witnesses has
been demonstrated for qubit states $\hat{\rho} \neq \textstyle{1
\over 2}\hat{I}$. In spite of that, the proposed scheme does not
give the explicit shape of the operators $\hat{A}$ and $\hat{B}$.
In this section, we are going to repair this gap.

First of all, the state $\hat{\rho}=\textstyle{1 \over 2}\hat{I}$
can be treated as classical, because its tomogram $w(m,u)\equiv
\textstyle{1 \over 2}$ and coincides with the case of classical
probabilities $p_{\uparrow}=p_{\downarrow}= \textstyle{1 \over
2}$. Let us demonstrate that all the other qubit states are
quantum.

Indeed, an arbitrary density operator $\hat{\rho}$ can be reduced
to the diagonal form of

\begin{equation}
\label{rho-diag}
\rho_{d} = \left(%
\begin{array}{cc}
  \frac{1+r}{2} & 0 \\
  0 & \frac{1-r}{2} \\
\end{array}%
\right)
\end{equation}

\noindent that becomes maximally mixed (classical) if $r=0$ and
pure if $r=1$. In order to find the quantumness witness for all
$0<r \le 1$ we avoid limitations ${\rm Tr}\hat{B}=1$ or ${\rm
Tr}\hat{B}=2$ imposed in \cite{alicki-small} and
\cite{alicki-large}, respectively. Below we introduce a family of
the operators $\hat{A}$ and $\hat{B}$ that satisfy the
requirements $0 \le \hat{A} \le \hat{B}$ while
$\hat{B}^{2}-\hat{A}^{2}$ is a quantumness witness for a
particular region of parameters $a$, $b$, $k$. For the diagonal
density matrix (\ref{rho-diag}) one can use

\begin{eqnarray}
&\hat{A}_{d} = \left(%
\begin{array}{cc}
  \frac{1}{2r^{2}}-\frac{a}{r} & \frac{1}{r}\sqrt{\frac{1}{4r^{2}}-a^{2}} \\
  \frac{1}{r}\sqrt{\frac{1}{4r^{2}}-a^{2}} & \frac{1}{2r^{2}}+\frac{a}{r} \\
\end{array}%
\right), \\ &\hat{B}_{d} = \hat{A}_{d} + \left(%
\begin{array}{cc}
  b r^{k} & -b r^{k} \\
  -b r^{k} & b r^{k} \\
\end{array}%
\right).
\end{eqnarray}

\begin{figure}
\begin{center}
\includegraphics{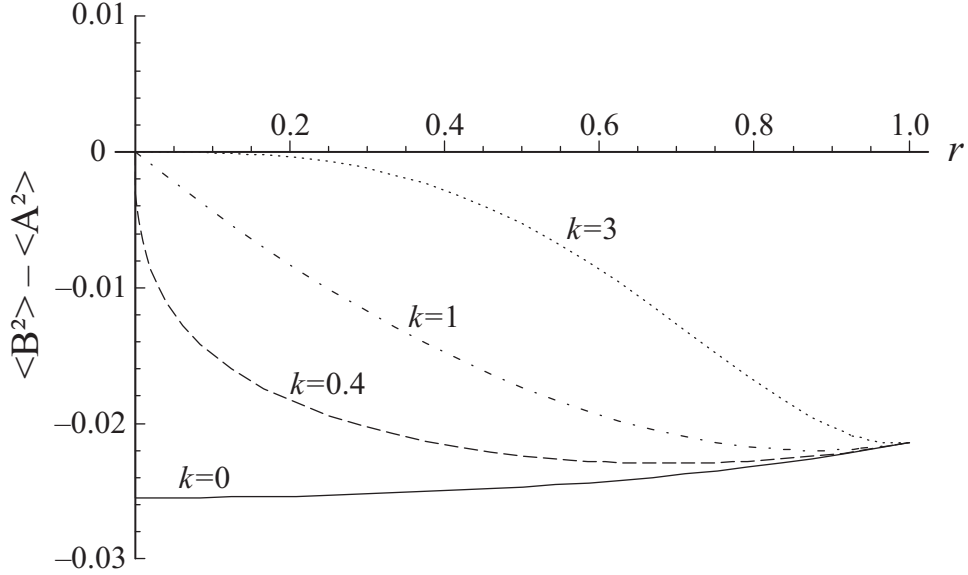}
\caption{\label{graph-1}The average value ${\rm
Tr}\left[\hat{\rho}_{d}(\hat{B}_{d}^{2}-\hat{A}_{d}^{2})\right]$
(expectation value of quantumness witness) is negative for all $r$
of the range $0< r \le 1$ ($a=0.35$, $b=0.1$, and the values of
$k$ are depicted in the figure).}
\end{center}
\end{figure}

\noindent The operators $\hat{A}_{d}$, $\hat{B}_{d}$, $\hat{B}_{d}
- \hat{A}_{d}$ have nonnegative eigenvalues only while the average
value ${\rm
Tr}\left[\hat{\rho}_{d}(\hat{B}_{d}^{2}-\hat{A}_{d}^{2})\right]$
is negative for all $r$ of the range $0<r \le 1$. Figure
\ref{graph-1} gives the illustration of this fact for particular
values of parameters $a$, $b$, $k$.

\bigskip

All-tomographic statement of the quantumness test can be
formulated as follows.

Suppose we have an experimentally measured spin tomogram $w(m,u)$
of a qubit state. The parameter $r$ of the diagonal density matrix
$\hat{\rho}_{d}$ can be found as

\begin{equation}
r = \max\limits_{u\in SU(2)} {(w_{\uparrow}(u)-w_{\downarrow}(u))}
= w_{\uparrow}(u_{\rho})-w_{\downarrow}(u_{\rho}).
\end{equation}

\noindent Then for the operators $\hat{A}$ and $\hat{B}$ leading
to the quantumness witness one can use

\begin{equation}
\hat{A} = u_{\rho}^{\dag}\hat{A}_{d}(r)u_{\rho}, \qquad \hat{B} =
u_{\rho}^{\dag}\hat{B}_{d}(r)u_{\rho}.
\end{equation}

\section{\label{conclusions}Conclusions}

To conclude we summarize the main results of the paper. We
demonstrated that the criterion of quantumness of a system state
found in \cite{alicki-small,alicki-large} can be formulated within
the framework of the probability representation of quantum states.
In this representation, the structure of the criterion is
clarified since it is formulated by means of inequalities, where
stochastic matrices involved (see (\ref{classical-inequalities})
and (\ref{violation})) have different properties. In case of
classical states the stochastic matrix providing inequality has
constant matrix elements. Conversely, in case of quantum states
the stochastic matrix providing analogous inequality has matrix
elements depending on unitary group elements. In view of this, for
some values of the unitary group elements the inequality can be
violated. One can extend the analysis of quantumness tests given
for qubits to systems with continuous variables, e.g., for photon
quadrature components.

\begin{acknowledgments}
V. I. M. thanks the Russian Foundation for Basic Research for
partial support under Project Nos. 07-02-00598 and 08-02-90300.
\end{acknowledgments}

\end{document}